# Synchronization in Scale-free Dynamical Networks: Robustness and Fragility

**Xiao Fan Wang**

*Automatic Control Laboratory, Department of Mechanical Engineering*

*University of Bristol, Bristol BS8 1TR United Kingdom*

<X.Wang@bristol.ac.uk>

**Guanrong Chen**

*Department of Electronic Engineering*

*City University of Hong Kong, Kowloon, Hong Kong*

<GChen@ee.cityu.edu.hk>

**Abstract**

Recently, it has been demonstrated that many large complex networks display a scale-free feature, that is, their connectivity distributions have the power-law form. In this paper, we investigate the synchronization phenomena in a scale-free dynamical network. We show that its synchronizability is robust against random removal of nodes, but is fragile to specific removal of the most highly connected nodes.

* Supported by the UK Engineering and Physical Sciences Research Council through the EPSRC grant number GR/M97923/01 and the Hong Kong GRC CERG Grant 9040565.

# I. INTRODUCTION

Recently, trying to understand the generic features that characterize the formation and topology of complex networks, much work has been devoted to the study of a large-scale complex system described by a network or a graph with complex topology, whose nodes are the elements of the system and whose edges represent the interactions among them. For example, the topology of the Internet can be studied at two different granularities. At the router level, each router is a node and two nodes are joined by a link when they are physically tied together. At the inter-domain level, each domain (subnetwork) is represented by a single node and each inter-domain interconnection is an edge [1]. In the case of the World Wide Web (WWW), the nodes are Web sites and they are joined when there is a hyper-link from one to the other [2-6]. Cell functioning is guaranteed by a complex metabolic network, in which the nodes are substrates and enzymes and the edges represent chemical interactions [7]. Even citations of scientific papers can be viewed as a network, where the nodes are papers and the edges correspond to citations among different papers [8]. The human society can be characterized by a huge social network, in which a node is an individual, an organization, or a country, connected by social interactions [9].

The apparent ubiquity of networks leads to a fascinating set of common problems concerning how network structure facilitates and constraints network behaviors. However, due to the large size and the complexity of interactions of such networks, it has become possible only very recently to gather and analyze the huge amount of data from such intricate systems due to the availability of modern computing power. It is expected that such topological information will be increasingly available and, thus, raise the possibility of understanding the topological and dynamical properties of very large-scale and complex networks in such a way that the study is based on real data, rather than just simulated and idealized information.

Traditionally, a network of complex topology is described by a random graph, for example, the graph of Erdos and Renyi (ER) [10-11]. Recently, in order to describe the transition from a regular network to a random network, Watts and Strogatz (WS) introduced the so-called small-world network [12-13]. A common feature of the ER model and the WS model is that the connectivity distribution of the network peaks at an average value and decays exponentially. Such an exponential network is homogeneous in nature: each node has roughly the same number of connections.

One significant recent discovery in the field of complex networks is the observation that a number of large-scale and complex networks are scale-free, that is, their connectivity distributions have the power-law form. This includes the above examples, as discussed in [14-15]. A scale-free network is inhomogeneous in nature: most nodes have very few connections and only a small number of nodes have many connections. It is this inhomogeneous feature that makes a scale-free network error tolerant but vulnerable to attacks [16-19]. More precisely, the connectivity of such networks is highly robust against random failures, that is, random removal of nodes (for example, random failures of routers in the Internet), but it is also extremely fragile to attacks, that is, specific removal of the most highly connected nodes. This may help explain why the Internet chugs right along despite frequent router failures. On the other hand, the average performance of the Internet would be cut in half if just 1% of the most highly connected routers were incapacitated. The Internet will even lose its integrity with 4% of its most important routers being destroyed. This is known as the '*Achilles heel*' of the Internet [16-17].

Connectivity is a topological property of networks. To better investigate the dynamical behaviors of complex networks, one may extend the existed scale-free network models by introducing dynamical elements into the network nodes. One of the most significant and interesting properties of a dynamical network is the synchronization motion of its dynamical elements. Synchronization has long been a basic concept in science and technology [20]. The ability of coupled oscillators to synchronize each other is a basis for the explanation of many processes of nature. In particular, synchronization in a network of coupled chaotic systems has recently become a topic of great interest [21-25]. This work is an effort devoted to the study of the synchronizability of a scale-free dynamical network, to show that it is robust against random removal of nodes, and yet is fragile to specific removal of the most highly connected nodes.

## II. THE SCALE-FREE DYNAMICAL NETWORK MODEL

*A. A Scale-Free Network Model*

Barabási and Albert suggested that two ingredients of self-organization of a network in a scale-free structure are growth and preferential attachment [14-15]. These refer to that networks continuously grow by the addition of new nodes and new nodes are preferentially attached to existing nodes with high numbers of connections (the so-called '*rich get richer*' phenomena).

In the present paper, a simple scale-free model of Barabasi and Albert is adopted. The model starts with $m_0$ nodes. At every time step, a new node is introduced, which is connected to $m$ already-existing nodes. The probability $\Pi_i$ that the new node is connected to node $i$ depends on the degree $k_i$ of node $i$ such that $\Pi_i = k_i / \sum_j k_j$. For large time, the probability $P(k)$ that a node in the network is connected to $k$ other nodes decays in a power-law of the form $P(k) = 2m^2 / k^3$ [14]. As a side note, Albert and Barabasi recently proposed an extended model of network evolution that gives a more realistic description of local processing, taking into account the additions of new nodes and new links, and the rewiring of links [26].

*B. The Scale-Free Dynamical Network Model*

Now suppose that at sometime, the scale-free network consists of $N$ identical linearly and diffusive coupled nodes, with each node being a $n$-dimensional dynamical system. The state equations of the network are

$$\dot{\mathbf{x}}_i = f(\mathbf{x}_i) + c \sum_{\substack{j=1 \\ j \neq i}}^{N} a_{ij} \Gamma (\mathbf{x}_j - \mathbf{x}_i) \quad i = 1, 2, \cdots, N \tag{1}$$

where $\mathbf{x}_i = (x_{i1}, x_{i2}, \cdots, x_{in}) \in \Re^n$ are the state variables of node $i$, the constant $c > 0$ represents the coupling strength and $\Gamma \in \Re^{n \times n}$ is a constant $0-1$ matrix linking coupled variables. For simplicity, we assume that $\Gamma = diag(r_1, r_2, \cdots, r_n)$ is a diagonal matrix with $r_i = 1$ for a particular $i$ and $r_j = 0$ for $j \neq i$. This means that two coupled nodes are linked through their $i$th state variables. If there is a connection between node $i$ and node $j$ ($i \neq j$), then $a_{ij} = a_{ji} = 1$; otherwise, $a_{ij} = a_{ji} = 0$ ($i \neq j$).

If the degree $k_i$ of node $i$ is defined to be the number of connection incidents on node $i$, then

$$\sum_{\substack{j=1 \\ j \neq i}}^{N} a_{ij} = \sum_{\substack{j=1 \\ j \neq i}}^{N} a_{ji} = k_i, \quad i = 1, 2, \cdots, N. \tag{2}$$

Let

$$a_{ii} = -k_i, \quad i = 1, 2, \cdots, N. \tag{3}$$

Then equations (1) can be written as

$$\dot{\mathbf{x}}_i = f(\mathbf{x}_i) + c \sum_{j=1}^{N} a_{ij} \Gamma \mathbf{x}_j, \quad i = 1, 2, \cdots, N \tag{4}$$

Coupling matrix $\mathbf{A} = (a_{ij}) \in \mathfrak{R}^{N \times N}$ represents the coupling configuration of the network. Suppose that the network is connected in the sense that there are no isolate clusters. Then the coupling matrix $\mathbf{A} = (a_{ij})_{N \times N}$ is a symmetric irreducible matrix. In this case, it can be shown that zero is an eigenvalue of $\mathbf{A}$ with multiplicity 1 and all the other eigenvalues of $\mathbf{A}$ are strictly negative.

The $i$th row $a(i,:)$ and $i$th column $a(:,i)$ of matrix $\mathbf{A}$ will be called as the $i$th row-column pair of $\mathbf{A}$. Equations (2) and (3) together implies that

$$\|a(i,:)\|_1 = \|a(:,i)\|_1 \equiv \sum_{j=1}^{N} |a_{ji}| = 2k_i \tag{5}$$

The power-law connectivity distribution makes a scale-free network extremely inhomogeneous: the majority of nodes are 'small' nodes with very small degrees, while a few nodes are 'big' nodes with very high degrees. This, in turn, implies that most row-column pairs of the coupling matrix have small 1-norms and a few row-column pairs have large 1-norms.

# III. SYNCHRONIZATION IN SCALE-FREE DYNAMICAL NETWORKS

*A. Synchronization Stability Analysis*

Let $\mathbf{s}(t) \in \Re^n$ be a solution of an isolate node, i.e.,

$$\dot{\mathbf{s}}(t) = f(\mathbf{s}(t)) \tag{6}$$

Here $\mathbf{s}(t)$ can be an equilibrium point, a periodic orbit or a chaotic attractor. Clearly, stability of the synchronization state

$$\mathbf{x}_1(t) = \mathbf{x}_2(t) = \cdots = \mathbf{x}_N(t) = \mathbf{s}(t) \tag{7}$$

of the network (4) is determined by the dynamics of an isolate node (i.e., function $f$ and solution $\mathbf{s}(t)$), the coupling strength $c$, the inner linking matrix $\Gamma$ and the coupling matrix $\mathbf{A}$.

Given the dynamics of an isolate node and the inner linking structure, the synchronizability of a network with respect to a specific coupling configuration is said to be *strong* if the network can synchronize with a small coupling strength. To investigate the stability of the synchronization state (7), we set

$$\mathbf{x}_i(t) = \mathbf{s}(t) + \boldsymbol{h}_i(t), \quad i = 1, 2, \cdots, N$$

and linearize Eq. (4) about $\mathbf{s}(t)$. This leads to

$$\dot{\boldsymbol{h}} = \boldsymbol{h}[Df(\mathbf{s})] + c\mathbf{A}\boldsymbol{h}\Gamma \tag{8}$$

where $\boldsymbol{h} = (\boldsymbol{h}_1, \boldsymbol{h}_2, \cdots, \boldsymbol{h}_N)^T \in \Re^{N \times n}$, $Df(\mathbf{s}) \in \Re^{n \times n}$ is the Jacobian of $f$ on $\mathbf{s}(t)$. Let

$$0 = l_1 > l_2 \geq l_3 \geq \cdots \geq l_N$$

be the eigenvalues of matrix $\mathbf{A}$ and $\Phi = [\boldsymbol{f}_1\ \boldsymbol{f}_2\ \cdots\ \boldsymbol{f}_N] \in \Re^{N \times N}$ be the corresponding (generalized) eigenvector basis such that

$$\mathbf{A}\boldsymbol{f}_k = l_k \boldsymbol{f}_k, \quad k = 1, 2, \cdots, N \tag{9}$$

By expanding each column $\boldsymbol{h}$ on the basis $\Phi$, we have

$$\boldsymbol{h} = \Phi \boldsymbol{u} \tag{10}$$

where the matrix $\boldsymbol{u} \in \Re^{N \times n}$ obey the following equations

$$\dot{\boldsymbol{u}} = \boldsymbol{u}[Df(s)] + c\Lambda\boldsymbol{u}\Gamma \tag{11}$$

where $\Lambda = diag(l_1, l_2, \cdots, l_N)$. Let $\boldsymbol{u}_k$ be the $k$th row of $\boldsymbol{u}$. We have

$$\dot{\boldsymbol{u}}_k^T = [Df(\boldsymbol{s}(t)) + cl_k \Gamma]\boldsymbol{u}_k^T, \quad k = 1, 2, \cdots, N \tag{12}$$

We have now transferred the stability problem of the synchronization state (7) to the stability problem of the $N$ $n$-dimensional linear time-varying systems (12). Note that $l_1 = 0$ corresponds to the synchronization state. If the following $N-1$ $n$-dimensional linear time-varying systems

$$\dot{w} = [Df(\boldsymbol{s}(t)) + cl_k \Gamma]w, \quad k = 2, \cdots, N \tag{13}$$

are exponentially stable, then $\boldsymbol{h}(t)$ will tends to the origin exponentially which implies that the synchronization state (7) is exponentially stable.

If $\boldsymbol{s}(t) = \bar{\boldsymbol{s}}$ is an equilibrium point, then a necessary and sufficient condition for the stability of systems (13) is that the real parts of the eigenvalues of the matrix $[Df(\bar{\boldsymbol{s}}) + cl_2 \Gamma]$ are all negative. Here we are particular interested in the case that $\boldsymbol{s}(t)$ is chaotic. A commonly used criterion for chaos synchronization is that all the transverse Lyapunov exponents of Eqs. (13) are negative [24]. However, it has been shown that this criterion is by no means a sufficient condition for synchronization. Intervals of desynchronized bursting behavior, called *attractor bubbling*, can appear even when the largest transverse Lyapunov exponent is negative [27]. Another method for the analysis of the stability of synchronization in chaotic systems is the Lyapunov function method, which is well-known in nonlinear systems theory [23]. Based on the Lyapunov stability theory, we have the following sufficient criterion for synchronization.

*Lemma 1:* Consider the network (4). Suppose that there exists a $n \times n$ diagonal matrix $\boldsymbol{D} > 0$, two constants $\bar{d} < 0$ and $t > 0$ such that

$$[Df(\boldsymbol{s}(t)) + d\Gamma]^T \boldsymbol{D} + \boldsymbol{D} [Df(\boldsymbol{s}(t)) + d\Gamma] \leq -t \boldsymbol{I}_n \tag{14}$$

for all $d \leq \bar{d}$, where $\boldsymbol{I}_n \in \mathfrak{R}^{n \times n}$ is an unit matrix. If

$$cl_2 \leq \bar{d} \tag{15}$$

*Then the synchronization state (7) is exponentially stable.*

*Proof:* Inequality (15) implies that

$$c l_k \leq \bar{d}, \quad k = 2, \cdots, N \tag{16}$$

From (14) and (16), we have

$$[Df(\mathbf{s}(t)) + c l_k \Gamma]^T \mathbf{D} + \mathbf{D}[Df(\mathbf{s}(t)) + c l_k \Gamma] \leq -t \mathbf{I}_n, \quad k = 2, \cdots, N \tag{17}$$

which means systems (13) are uniformly exponentially stable with Lyapunov functions

$$V_k = w^T \mathbf{D} w, \quad k = 2, \cdots, N. \tag{18}$$

Since $l_2 < 0$ and $\bar{d} < 0$, inequality (15) is equivalent to

$$c \geq \left| \frac{\bar{d}}{l_2} \right| \tag{19}$$

A small value of $l_2$ corresponds to a large value of $|l_2|$ which implies that the network (4) can synchronize with a small coupling strength $c$. Therefore, synchronizability of network (4) with respect to a specific coupling configuration can be characterized by the second-largest eigenvalue $l_2$ of the corresponding coupling matrix $\mathbf{A}$.

*B. Synchronizability of the Scale-Free Dynamical Network*

For clarity, we use $\mathbf{A}_{sf}(\bar{m}, N)$ to denote the coupling matrix of a scale-free dynamical network (4), which has $N$ nodes and $\bar{m}(N - \bar{m} - 1)$ connections (i.e., $m_0 = m = \bar{m}$). Let $l_{2sf}(\bar{m}, N)$ be the second-largest eigenvalue of $\mathbf{A}_{sf}(\bar{m}, N)$. Numerical computation reveals that for a fixed value of $\bar{m}$, $l_{2sf}(\bar{m}, N)$ increases to a negative constant $l_{2sf}(\bar{m})$ as $N$ increases. Here for each pair of values of $\bar{m}$ and $N$, $l_{2sf}(\bar{m}, N)$ is obtained by averaging the results of 20 runs. In particular, for $\bar{m} = 3, 5, 7, 9$ and $11$, one has $l_{2sf}(\bar{m}) \approx -0.944, -0.973, -0.981, -0.983$ and $-0.985$, respectively, as shown in Fig. 1.

Therefore, the adding of new nodes in a scale-free network can not decrease the synchronizability of the network. In fact, due to the self-organization process a scale-free network, the synchronizability of a large size scale-free dynamical network will remain almost unchanged by the constantly adding of new nodes. The numerical results also show that the value of $\overline{m}$ (i.e., the number of connections when a new node is connected to the existed nodes) has only a minor influence on the synchronizability of a scale-free network. In the thermodynamic limit case, we have the following conjecture.

*Conjecture 1:* For any given constant $\overline{m} > 1$, we have

$$\lim_{N \to \infty} \lambda_{2sf}(\overline{m}, N) = \hat{\lambda}_{2sf} \tag{20}$$

where $\hat{\lambda}_{2sf}$ is a constant that is unrelated with $\overline{m}$.

*C. Compare to the Locally Regular Coupled Network*

Over the past decade, much work on synchronization in coupled oscillator arrays had been emphasised on regular coupling scheme. One typical case is a locally coupled regular network with periodic boundary condition $\mathbf{x}_{N+j} = \mathbf{x}_j$ in which each node $i$ is adjacent to its neighbour nodes $i \pm 1$, $i \pm 2$, …, $i \pm l$, $l$ is a positive integer. The state equations of the network are

$$\dot{\mathbf{x}}_i = f(\mathbf{x}_i) + c \sum_{j=1}^{l} \Gamma(\mathbf{x}_{i+j} + \mathbf{x}_{i-j} - 2\mathbf{x}_i), \ i = 1, 2, \cdots, N \tag{21}$$

The coupling matrix $\mathbf{A}_{lc}$ of network (21) is a circulant matrix and its second-largest eigenvalue can be computed as

$$\lambda_{2lc} = -4 \sum_{j=1}^{l} \sin^2 \left( \frac{j\pi}{N} \right) \tag{22}$$

The regular network (21) contains $N$ nodes and $(2l-1)N$ connections. Let $\overline{m} = 2l - 1$. For large $N$, the regular network (21) and the scale-free network (4) have about the same number of connections. However, the synchronizability of the two networks is quite different: as the number of nodes increases to infinity, the second-largest eigenvalue of the locally coupled regular network (21) decreases to zero, while the second-largest eigenvalue of the scale-free network (4) increases to a negative constant. Therefore, it's almost practically impossible to achieve synchronization in a very large network with only locally coupled configuration.

*D. An Example*

Now we illustrate the above analysis using a Chua's oscillator as a dynamical node. In the dimensionless form, a single Chua's oscillator is described by [28]:

$$\begin{pmatrix} \dot{x}_1 \\ \dot{x}_2 \\ \dot{x}_3 \end{pmatrix} = \begin{pmatrix} \mathbf{a}(x_2 - x_1 + f(x_1)) \\ x_1 - x_2 + x_3 \\ -\mathbf{b}x_2 - \mathbf{g}x_3 \end{pmatrix}, \qquad (23)$$

where $f(.)$ is a piecewise linear function,

$$f(x_1) = \begin{cases} -bx_1 - a + b & x_1 > 1 \\ -ax_1 & |x_1| \leq 1 \\ -bx_1 + a - b & x_1 < -1 \end{cases} \qquad (24)$$

in which $\mathbf{a} > 0, \mathbf{b} > 0, \mathbf{g} > 0$, and $a < b < 0$. Suppose that two coupled Chua's oscillators are linked through the first state variable, i.e., $\Gamma = diag(1, 0, 0)$. The state equations of the entire network are

$$\begin{pmatrix} \dot{x}_{i1} \\ \dot{x}_{i2} \\ \dot{x}_{i3} \end{pmatrix} = \begin{pmatrix} \mathbf{a}(x_{i2} - x_{i1} + f(x_{i1})) + c \sum_{j=1}^{N} a_{ij} x_{j1} \\ x_{i1} - x_{i2} + x_{i3} \\ -\mathbf{b}x_{i2} - \mathbf{g}x_{i3} \end{pmatrix}, \; i = 1, 2, \cdots, N. \qquad (25)$$

For network (25), it can be checked that the constant $\overline{d}$ in Lemma 1 can be taken as $\overline{d} = -a$. If the system parameters are chosen to be

$$\mathbf{a} = 10.0000, \; \mathbf{b} = 15.0000, \; \mathbf{g} = 0.0385, \; a = -1.2700, \; b = -0.6800, \qquad (26)$$

then Chua's oscillator (23) has a chaotic attractor, as shown in Fig 2. For sufficiently large number of nodes, chaotic synchronization of the network (25) with a scale-free coupling structure can be achieved, provided that

$$c > \left| \frac{a}{\overline{l}_{2sf}(\overline{m})} \right|. \qquad (27)$$

For instance, for $\overline{m} = 3, 5$ and $7$, synchronization can be achieved with $c > 1.346$, $c > 1.306$ and $c > 1.295$, respectively.

### III. ROBUSTNESS AND FRAGILITY OF SYNCHRONIZATION

Now we consider the robustness of synchronizability in scale-free dynamical networks against either randomly or specifically removal of a small fraction $f$ $(0 < f << 1)$ of nodes in the network. Clearly, the removal of some nodes in a network (4) can only change the coupling matrix. According to the above analysis, it amounts to studying the changes in the second-largest eigenvalue of the coupling matrix of the network. If the second-largest eigenvalue of the coupling matrix remains unchanged, then the synchronization stability of the network will also remain unchanged after the removal of some nodes.

For simplicity, we use $\mathbf{A}_{sf} \in \mathfrak{R}^{N \times N}$ and $\tilde{\mathbf{A}}_{sf} \in \mathfrak{R}^{(N-[fN]) \times (N-[fN])}$ to denote the coupling matrix of the original network with $N$ nodes and the network after removal of $[fN]$ nodes, respectively. Here, $[fN]$ stands for the smaller but nearest integer to the real number $fN$. Let $l_{2sf}$ and $\tilde{l}_{2sf}$ be the second-largest eigenvalues of $\mathbf{A}_{sf}$ and $\tilde{\mathbf{A}}_{sf}$, respectively. In the following simulations, $N = 3000$ and $m_0 = m = 3$, namely, the original network contains 3000 nodes and about 9000 connections.

*A. Robustness of Synchronization against Failures*

Suppose that nodes $i_1, i_2, ..., i_{[fN]}$ have been removed from the network. One can construct the new coupling matrix $\tilde{\mathbf{A}}_{sf}$ from the original coupling matrix $\mathbf{A}_{sf}$ as following: First, form the minor matrix $\mathbf{B} \in \mathfrak{R}^{(N-[fN]) \times (N-[fN])}$ of $\mathbf{A}_{sf}$ by removing the $i_1 th, i_2 th, ..., i_{[fN]} th$ row-column pairs of $\mathbf{A}_{sf}$. Then, obtain $\tilde{\mathbf{A}}_{sf}$ by re-computing the diagonal elements of the above minor matrix according to formulas (2)-(3) with $N$ being replaced by $(1-f)N$.

It was found that even when as many as 5% of randomly chosen nodes are removed, the second-largest eigenvalue of the coupling matrix remains almost unchanged (see Fig. 3), i.e.,

$$\tilde{l}_{2sf} \approx l_{2sf} . \tag{28}$$

This implies that the synchronizability of the network is almost unaffected.

The error tolerance of synchronizability in scale-free networks may be due to their extremely inhomogeneous connectivity distributions. Since most of the nodes in a scale-free network are 'small' nodes with very low degrees, $[fN]$ 'small' nodes will be selected with much higher probability if $N >> 1$ and $f << 1$. The removal of these 'small' nodes does not alter the path structure of the remaining nodes and, thus, does not destroy the connectivity of the network. This means that the coupling matrix $\tilde{\mathbf{A}}_{sf}$ remains to be an irreducible and symmetric matrix. Furthermore, a 'small' node with a low degree corresponds to a row-column pair with small 1-norm. So it is reasonable to conjecture that the second-largest eigenvalue of the coupling matrix $\mathbf{A}_{sf}$ remains almost unchanged by taking such row-column pairs with small 1-norms. That is, if the removed nodes $i_1, i_2,..., i_{[fN]}$ are all 'small' nodes with very low degrees, then one should have

$$l_2(\mathbf{B}) \approx l_{2sf} \tag{29}$$

where $l_2(\mathbf{B})$ is the second-largest eigenvalue of the minor matrix $\mathbf{B}$.

On the other hand, from the construction of the new coupling matrix $\tilde{\mathbf{A}}_{sf}$, it is easy to see that

$$\tilde{\mathbf{A}}_{sf} = \mathbf{B} + \mathbf{D} , \tag{30}$$

where $\mathbf{D} = diag(d_{11}, d_{12}, \cdots, d_{N-[fN]})$ is a nonnegative diagonal matrix with

$$\sum_{j=1}^{N-[fN]} d_{jj} = \sum_{j=1}^{[fN]} k_{i_j} . \tag{31}$$

If the right-hand side of (31) is sufficiently small, one may deduce that

$$\tilde{l}_{2sf} \approx l_2(\mathbf{B}). \tag{32}$$

(29) and (32) together imply that (28) holds, i.e., the second-largest eigenvalue of the coupling matrix remains almost unchanged after the removal of a small fraction of randomly chosen nodes.

Although the scale-free structure is particularly well-suited to tolerate random errors, it is also particularly vulnerable to deliberate attacks.

To simulate the influence of an attack on the synchronizability of the network, one may first remove the node with highest degree, and then continue to select and remove other nodes in decreasing order of degrees. In doing so, it was found in Fig. 4 that the second-largest eigenvalue of the coupling matrix increases rapidly, almost decreases to half of its original value in magnitude (from the original $\lambda_{2sf} = -0.953$ to $\tilde{\lambda}_{2sf} = -0.519$), when only $f = 1\%$ fraction of the most connected nodes was removed. At a low critical threshold, $f = 1.6\%$, $\tilde{\lambda}_{2sf}$ abruptly changes to zero, implying that the whole network was broken into isolate clusters.

This vulnerability of connectivity and synchronizability to attack in scale-free networks is also rooted in their extremely inhomogeneous connectivity distribution. The network dynamics are dominated by a small number of 'big' nodes with high degrees. The removal of a small fraction of such 'big' nodes implies the minor matrix $\mathbf{B} \in \mathfrak{R}^{(N-[fN]) \times (N-[fN])}$ of $\mathbf{A}_{sf}$ is derived by removing a small fraction of row-column pairs of $\mathbf{A}_{sf}$ with large 1-norms. However, we find that the second-largest eigenvalue of $\mathbf{B}$ approximately equal to that of the original matrix $\mathbf{A}_{sf}$, i.e., we still have

$$\lambda_2(\mathbf{B}) \approx \lambda_{2sf}$$

On the other hand, in this case, we can see from (30) and (31) that the new coupling matrix $\tilde{\mathbf{A}}_{sf}$ is derived from matrix $\mathbf{B}$ by adding a nonnegative diagonal matrix with a large sum of the diagonal elements. This would results in a drastic change in the second-largest eigenvalue of the matrix, i.e., $\tilde{\lambda}_{2sf}$ is much greater than $\lambda_2(\mathbf{B})$ and $\lambda_{2sf}$.

The above analysis, along with supportive numerical results, suggests the following conjecture:

*Conjecture 2:* Suppose that $N \gg 1$, $0 \leq f \ll 1$. Let $\mathbf{B}$ be a minor matrix of $\mathbf{A}_{sf}$ by arbitrarily removing $[fN]$ row-column pairs of $\mathbf{A}_{sf}$. Then

$$\lambda_2(\mathbf{B}) \approx \lambda_{2sf}$$

Moreover, if $[fN]$ sufficiently small nodes are deleted, i.e., $\{k_{i_j}\}_{j=1}^{[fN]}$ are sufficiently small, then

$$\tilde{\lambda}_{2sf} \approx \lambda_2(\mathbf{B}) \approx \lambda_{2sf} \tag{33}$$

## V. CONCLUSIONS

The work of Albert *et. al.* [16] and the work presented in this paper have shown that both the connectivity and the synchronizability of a scale-free dynamical network are robust against random removal of nodes but fragile to some specific removal of nodes. These results, together with some other recent findings about the robustness and power-law in complex systems (for example, see [27-29]), indicate that 'robust yet fragile' seems to be a generic feature for topological and dynamical properties of scale-free networks that obey power-law distributions.

# REFERENCES

[1] M. Faloutsos, P. Faloutsos and C. Faloutsos, "On power-law relationships of the Internet topology," *Comput. Commun. Rev.,* vol. **29**, pp. 251-263, 1999.

[2] R. Albert, H. Jeong and A.-L. Barabási, "Diameter of the World Wide Web," *Nature,* vol. **401**, pp. 130-131, 1999.

[3] A.-L. Barabási, R. Albert and H. Jeong, "Scale-free characteristics of random networks: The topology of the world-wide web," *Physica A*, vol. **281**, pp. 2115, 2000.

[4] A.-L. Barabási, R. Albert, H. Jeong and G. Bianconi, "Power-law distribution of the World Wide Web," *Science,* vol. **287**, pp. 2115a, 2000.

[5] A. Broder, Kumar, F. Maghoul, P. Raghavan, S. Rajagopalan, R. Stata, A. Tomkins, and J. Wiener, "Graph structure in the Web," *Computer Networks,* vol. **33**, pp. 309-320, 2000.

[6] B. A. Huberman, and L. A. Adamic, "Growth dynamics of the World-Wide Web," *Nature,* vol. **401**, pp. 131, 1999.

[7] H. Jeong, B. Tombor, R. Albert, Z. Oltvai, and A.-L. Barabási, "The large-scale organization of metabolic networks," *Nature,* vol. **407**, pp. 651-653, 2000.

[8] S. Redner, "How popular is your paper? An empirical study of the citation distribution," *Eur. Phys. J. B,* vol. **4,** no. 2, pp. 131-134, 1998.

[9] S. Wassrman, and K. Faust, *Social Network Analysis*, Cambridge University Press, Cambridge, 1994.

[10] P. Erdos, and A. Renyi, "On the evolution of random graphs," *Publ. Math. Inst. Hung. Acad. Sci.*, vol. 5, pp. 17-60, 1960.

[11] B. Bollobas, *Random Graphs*, Academic, London, 1985.

[12] D. J. Watts, and S. H. Strogatz, "Collective dynamics of 'small world' networks," *Nature,* vol. **393**, pp. 440-442, 1998.

[13] D. J. Watts, *Small Worlds: The Dynamics of Networks between Order and Randomness.* Princeton University Press, Princeton, 1999.

[14] A.-L. Barabási and R. Albert, "Emergence of scaling in random networks," *Science, vol.* **286**, pp. 509-512, 1999.

[15] A.-L. Barabási, R. Albert and H. Jeong, "Mean-field theory for scale-free random networks," *Physica*, vol. **272**, pp. 173-187, 1999.

[16] R. Albert, H. Jeong and A.-L. Barabási, "Attack and error tolerance in complex networks," *Nature*, vol. **406**, pp. 387-482, 2000.

[17] Y. Tu, "How robust is the Internet?" *Nature,* vol. **406**, pp. 353-354, 2000.

[18] R. Cohen, K. Erez, D. Ben-Avraham, and S. Havlin, "Resilience of the Internet to random breakdowns," *Phys. Rev. Lett.*, vol. **85**, no. 21, 4626-4628, 2000.

[19] D. S. Callway, M. E. J. Newman, S. H. Strogatz, and D. J. Watts, "Network robustness and fragility: Percolation on random graphs," *Phys. Rev. Lett.*, vol. 85, no. 25, 5468-5471, 2000.

[20] I. I. Blekhman, *Synchronization in Science and Technology*. English translation: New York: ASME, 1988.

[21] L. M. Pecora, and T. L. Carroll, "Synchronization in chaotic systems," *Phys. Rev. Lett.*, vol. **64,** no. 8, pp. 821-824, 1990.

[22] J. F. Heagy, T. L. Carroll, and L. M. Pecora, "Synchronous chaos in coupled oscillator systems," *Phys. Rev. E,* vol. **50,** no. 3, pp. 1874-1885, 1994.

[23] C. W. Wu, and L. O. Chua, "Synchronization in an array of linearly coupled dynamical systems," *IEEE Trans. Circuits Syst.-I*, vol. 42, no. 8, pp. 430-447, 1995.

[24] G. Hu, J. Yang, and W. Liu, "Instability and controllability of linearly coupled oscillators: Eigenvalue analysis," *Phys. Rev. E*, vol. 58, no. 4, pp. 4440-4453, 1998.

[25] L. M. Pecora, T. L. Carroll, G. Johnson, D. Mar, and K. S. Fink, "Synchronization stability in coupled oscillator arrays: Solution for arbitrary configurations," Int. J. Bifurcation Chaos, vol. 10, no. 2, pp. 273-290, 2000.

[26] R. Albert and A.-L. Barabási, "Topology of complex networks: Local events and universality," *Phys. Rev. Lett.*, vol. **85,** no. 24, pp. 5234-5237, 2000.

[27] Sushchik, M. Mikhail, Rulkov, F. Nikolai, and H. D. I. Abarbanel, "Robustness and stability of synchronized chaos: An illustrative model," *IEEE Trans. on Circuits and Systems-I*, vol. 44, no. 10, pp. 867-873, 1997.

[28] L. O. Chua, C. W. Wu, A. Huang, and G. Q. Zhong, "A universal circuit for studying and generating chaos, Part I+II," *IEEE Trans. Circuits and Systems-I,* vol. **40,** no. 10, 732-761, 1993.

[29] J. Carlson, and J. Doyle, "Highly optimized tolerance: A mechanism for power laws in designed systems," *Phys. Rev. E,* vol. **60**, no.2, 1412-1427, 1999.

[30] J. Carlson, and J. Doyle, "Highly optimized tolerance: Robustness and power laws in complex systems," *Phys. Rev. Lett.,* vol. **84**, no. 11, pp. 2529-2532, 2000.

[31] J. Doyle, and J. Carlson, "Power laws, highly optimized tolerance and generalized source coding," *Phys. Rev. Lett,.*vol. **84**, no. 24, pp. 5656-5659, 2000.

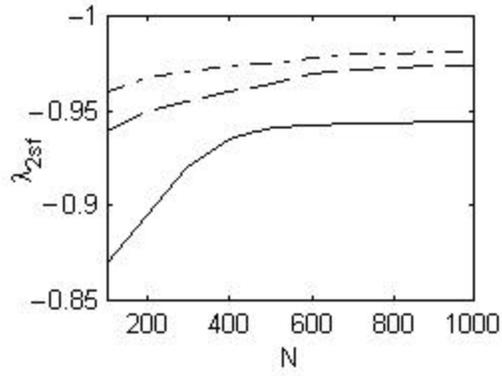

**Fig. 1** The second-largest eigenvalue of the coupling matrix of the scale-free network (4) for $m_0 = m = 3$ (—); $m_0 = m = 5$ (– –); and $m_0 = m = 7$ (– ·).

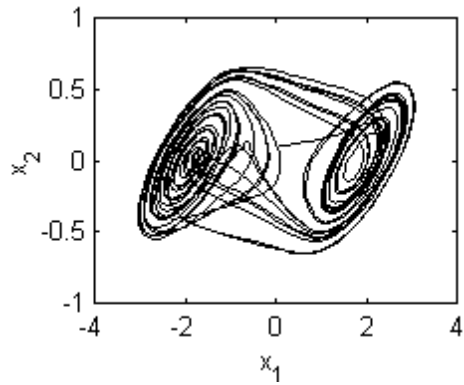

**Fig. 2** Chaotic attractor of the Chua's oscillator (23), with parameters given in (26).

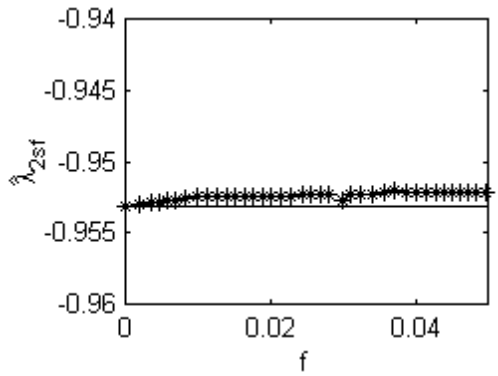

**Fig. 3** Changes in the second-largest eigenvalue of the coupling matrix of the scale-free network (4) when a fraction $f$ of the randomly selected nodes is removed.

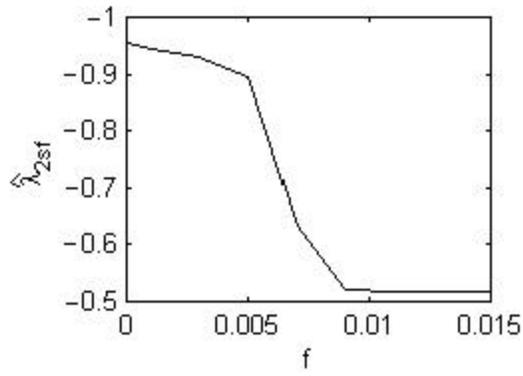

**Fig. 4** Changes in the second-largest eigenvalue of the coupling matrix of the scale-free network (4) when a fraction $f$ of the most connected nodes is removed.